\documentclass[11pt]{article}

\pdfoutput=1

\usepackage{authblk}
\usepackage{comment}
\usepackage{geometry}
\usepackage{colortbl}
\usepackage{graphicx}
\usepackage{makecell}
\usepackage{hyperref}
\usepackage{listings}

\begin{document}

\title{Database Traffic Interception \\ for Graybox Detection of Stored and Context-Sensitive XSS}

\author{Anton\'in Steinhauser}
\author{Petr T\r{u}ma}
\affil{%
  Department of Distributed and Dependable Systems
  
  Faculty of Mathematics and Physics, Charles University

  Malostransk\'e n\'am\v{e}st\'i 25, Prague, Czech Republic

  \texttt{\{name.surname\}@d3s.mff.cuni.cz}
}

\maketitle

\begin{abstract}
XSS (cross site scripting) is a type of a security vulnerability that permits injecting malicious code into the client side of a web application. In the simplest situations, XSS vulnerabilities arise when a web application includes the user input in the web output without due sanitization. Such simple XSS vulnerabilities can be detected fairly reliably with blackbox scanners, which inject malicious payload into sensitive parts of HTTP requests and look for the reflected values in the web output.

Contemporary blackbox scanners are not effective against stored XSS vulnerabilities, where the malicious payload in an HTTP response originates from the database storage of the web application, rather than from the associated HTTP request. Similarly, many blackbox scanners do not systematically handle context-sensitive XSS vulnerabilities, where the user input is included in the web output after a transformation that prevents the scanner from recognizing the original value, but does not sanitize the value sufficiently. Among the combination of two basic data sources (stored vs reflected) and two basic vulnerability patterns (context sensitive vs not so), only one is therefore tested systematically by state-of-the-art blackbox scanners.

Our work focuses on systematic coverage of the three remaining combinations. We present a graybox mechanism that extends a general purpose database to cooperate with our XSS scanner, reporting and injecting the test inputs at the boundary between the database and the web application. Furthermore, we design a mechanism for  identifying the injected inputs in the web output even after encoding by the web application, and check whether the encoding sanitizes the injected inputs correctly in the respective browser context. We evaluate our approach on eight mature and technologically diverse web applications, discovering previously unknown and exploitable XSS flaws in each of those applications.
\end{abstract}

\section* {Rights}

\noindent Uploaded to ArXiV under the ACM Copyright Policy Version 9.
Copyright 2020 ACM and the authors.
This is the author version of the work.
Posted for your personal use.
Not for redistribution.
The definitive Version of Record was published in
\emph{Digital Threats: Research and Practice (DTRAP)}
Vol. 1 No. 3, \url{https://doi.org/10.1145/3399668}.

\section{Introduction}

\label{sec:intro}

XSS (cross site scripting) is a security flaw particular to web applications. XSS flaws arise when a web application includes malicious input in the web output without due sanitization. When such output is processed by the victim browser, the malicious input can change the application behavior, resulting in a security attack. XSS flaws may permit a range of attack options, with the most severe being the ability to execute arbitrary JavaScript inside the victim browser. This can in turn lead to serious security incidents such as account hijacking.\footnote{The Real Impact of Cross-Site Scripting: \\ \url{https://www.dionach.com/blog/the-real-impact-of-cross-site-scripting}} A recent report\footnote{The Hacker-Powered Security Report 2018: \\ \url{https://www.hackerone.com/resources/hacker-powered-security-report}} lists XSS as the most common vulnerability across most surveyed industries.

To detect XSS vulnerabilities, an analyzer may adopt a whitebox or a blackbox approach. A whitebox XSS analyzer can be employed when it is possible to track the flow of potentially malicious inputs through the internals of the application~\cite{Huang:2004:SWA:988672.988679, jovanovic2006pixy, livshits2005finding, Tripp:2009, haldar2005dynamic, Steinhauser:2016:JSD:2993600.2993606, doi:10.1002/spe.2649}. In contrast, a blackbox XSS analyzer uses only external application interfaces, looking for injected input in application output. As an advantage, blackbox analysis can remain largely independent from the technology stack of the analyzed application~\cite{srinivas_nidhra_2012_1278008}, however, limiting the analyzer to the external application interfaces typically brings lower recall~\cite{6113264}.

In the last decade, blackbox XSS analyzers were shown to be ineffective against stored XSS flaws~\cite{10.1007/978-3-642-14215-4_7,5504795,7412085}. Compared to reflected XSS flaws~\footnote{Stored and reflected XSS attacks, OWASP: \\
\url{https://www.owasp.org/index.php/Cross-site\_Scripting\_(XSS)\#Stored\_and\_Reflected\_XSS\_Attacks}}, which return the malicious input in the response to the very request that submitted it, stored XSS flaws trick the application into recording the malicious input in persistent storage and returning it with some later response. Persistent input passes through increasingly complex user interfaces with captchas, CSRF protection tokens, consistency checks and other mechanisms that make it difficult for a blackbox XSS analyzer to successfully inject the test payload.

\medskip

\begin{figure*}[!ht]
\centering
\includegraphics{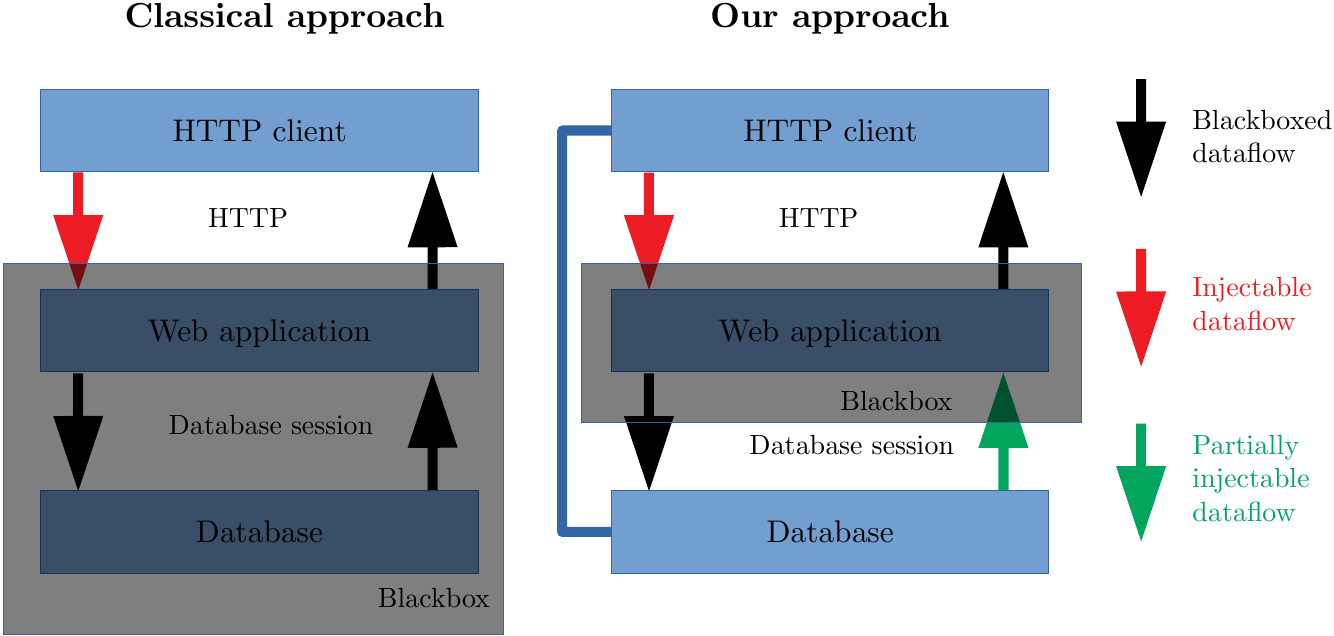}
\caption{Comparison of our graybox XSS analyzer architecture with classical blackbox XSS analyzer architecture.}
\label{fig:architecture}
\end{figure*}

Our work is based on observing that most persistent user input ends up in a database, which is typically a well-defined application component with a standard interface. An XSS analyzer can use this interface for injecting the test payload while still remaining reasonably application independent -- the database component is often connected through a socket that uses a well-defined communication protocol, hence intercepting the communication between the database and the application is technically close to intercepting the external application communication. For a comparison of our graybox analyzer architecture with a common blackbox XSS analyzer architecture, see Figure~\ref{fig:architecture}.

Strictly speaking, our approach is positioned between classical blackbox and whitebox analyses, as it requires the web application to use a specific client-server database protocol. However, it does not perform a classical graybox analysis, because we never access or modify the code that we analyze~\cite{BenJaballah:2016:GAD:2995959.2995966,sawant2012software,aitel2002advantages}. In the context of web application testing, where the web applications are usually tested independently of the database implementations, our approach is closer to blackbox analysis than to whitebox analysis.

Our graybox XSS analyzer combines classical HTTP parameter injection with injection to values fetched from the database by the web application. For this purpose, we have designed a communication protocol between the analyzer and the database and developed a prototype implementation of a database that supports this protocol and injects values provided by the analyzer into the web application. We identify the injected values in the HTML output using regular expression matching, taking care to also recognize values that were encoded by the web application. Afterwards, we parse the HTML output in order to compute the browser contexts~\cite{Steinhauser:2016:JSD:2993600.2993606, doi:10.1002/spe.2649, Samuel:2011:CAW:2046707.2046775} of the injected values, and check whether the applied sanitizations are compatible with those contexts. If not, we report an XSS vulnerability.

Our contributions are:

\textbf{Database Traffic Interception.} We introduce a mechanism through which the XSS analyzer can cooperate with the database used by the web application under analysis. The database can inform the analyzer about the values fetched by the application, and the analyzer can instruct the database to inject a value into a specific database fetch. The value can later be tracked down in the HTML output.

\textbf{Exploit Payload Identification.} We describe how we use regular expression matching to find the injected payload in the HTML output even when possibly encoded by the web application.

\textbf{Browser Context Computation.} We explain how we recursively parse the HTML output with our simplified web browser model in order to determine the browser context of the injected payload, and how we verify whether the encoding applied to the payload was sufficient for thus determined context.

\textbf{Evaluation with XSS Flaw Discoveries.} We demonstrate the usefulness of our XSS analyzer prototype by analyzing eight mature applications written in five different programming languages and serving diverse security-sensitive purposes. In addition to reporting correct and incorrect sanitization counts, we also examine whether and how the discovered bugs can be exploited, provide performance metrics for our analysis and compare the results with three popular state-of-the-art blackbox XSS scanners. Our analyzer discovered previously unknown exploitable XSS vulnerabilities in each of the analyzed applications, with the count and severity of the flaws differing significantly between the analyzed applications.

\medskip

In Section~\ref{sec:persistent} we describe the stored XSS flaws in more detail and discuss the complications that arise when we use database traffic interception for XSS flaw detection. In Section~\ref{sec:example} we explain the context-sensitive XSS flaws and their relevance. Section~\ref{sec:approach} describes in more detail how our graybox XSS analyzer works and how it cooperates with the database. In Section~\ref{sec:eval} we evaluate our approach on eight open source applications and discuss the evaluation results. Finally, Section~\ref{sec:relwork} discusses related work, and Section~\ref{sec:conclusion} closes the paper with concluding remarks and notes on future work.

\section{Stored XSS Flaws and Database Traffic Interception}
\label{sec:persistent}

There are two basic XSS flaw types -- reflected (or non-persistent) and stored (or persistent)~\footnote{Types of Cross-Site Scripting: \url{https://www.owasp.org/index.php/Types_of_Cross-Site_Scripting}}. Reflected XSS returns the unsafe client input immediately, in the HTTP response to the HTTP request that submitted the input, and therefore to the same client. A stored XSS flaw exists when the unsafe client input is stored by the HTTP server in a persistent storage and returned later, in another HTTP response to (possibly) another client.

The detection of stored XSS flaws is technically more complicated than the detection of reflected XSS flaws, because the stored XSS flaws have a longer and more complicated dataflow path for the exploit payload, and the malicious input can persist in the storage over many HTTP requests. By intercepting the database fetches and injecting test payload, we can make the stored XSS flaw detection conceptually as simple as the reflected XSS flaw detection, because we shorten the critical dataflow path to roughly the same length and complexity. However, this technique brings three drawbacks related to analysis precision, discussed below.

\medskip

The first drawback rests with the fact that the database traffic interception interface may offer payload injection opportunities that do not exist at the web application interface. Potential XSS flaws detected by database traffic interception therefore require verifying the existence of the dataflow path from the web application interface to the database, which the analysis bypasses. This verification may naturally lead to a negative verdict, for example when the database stores a value that is hardcoded in the application and therefore cannot be modified through the web interface, or when the value is sanitized by the web application before it is stored in the database.

Common security strategies usually do not assume that the attacker has full write access to the database. This may create the impression that using hardcoded or previously sanitized values from the database without additional checks is fine, and that our analysis will inherently report a wide class of false positives not exploitable using standard web application access methods. Our results demonstrate there are still good reasons to consider all data coming from the database unsafe:

\begin{itemize}
\item The sanitization correctness depends on how the database value will be used. The code that stores the value usually cannot easily deduce all eventual uses of the value, especially when those uses are implemented in independent application components.

\item Maintaining the safety of the database values over time represents an extra operational concern. In particular, later extensions to the application can make originally hardcoded values modifiable.

\item Relying on database value safety may amplify otherwise non-exploitable security flaws, such as an SQL injection into an INSERT INTO or an UPDATE statement.

\item Finally, the code-data separation paradigm suggests that even full write access to the database should not grant the ability to execute arbitrary JavaScript code in the client browsers.
\end{itemize}

\medskip

The second drawback of database traffic interception is that the injected values can be unrealistic,
fail to meet some consistency criteria or other expectations of the web application,
and even make the web application crash
(to a lesser degree, HTTP request parameter injection is affected by the same issue).
In a reasonably designed web application such an error should typically impact only the processing of that particular HTTP request,
leading to the 500 HTTP response code or similar. Such a failure can lead to a false negative but is not in principle prohibitive.

The issue may be exacerbated for those types of values that are more likely to affect application control flow, such as configuration information.
Section~\ref{subsec:payload} details one of the performance optimizations implemented in our approach, which
restricts injection to those database fields whose content was previously observed in the web application output.
Configuration information and similar types of values are likely outside this category,
hence our optimization may also provide partial remedy on this point.

\medskip

The third drawback concerns potential caching of either database or application responses.
The cache keys are usually derived from the request content. When the analysis injects
the exploit payload into requests, it obviously changes the cache keys and does not
use the previously cached values. However, response injection at database traffic level
may leave the cache keys unchanged, leading to false alarms.
Our XSS flaw analysis should therefore be done with caching disabled.

\section{Context-Sensitive XSS Flaws}
\label{sec:example}

Web applications defend against XSS flaws by sanitizing potentially malicious input. Importantly, sanitization is not a one-size-fits-all operation, but instead must consider where the input will be used. Failure to do so may lead to context-sensitive XSS flaws, which were shown to exist even in widely used web applications~\cite{Steinhauser:2016:JSD:2993600.2993606, doi:10.1002/spe.2649}.
Throughout the paper, we use the following running example, which illustrates context sensitivity on this condensed JavaScript and PHP snippet:

\begin{lstlisting}
Topic: <input id="topic" name="topic" />
<script>
function populateTopic(value) {
  par = document.getElementById("topic");
  par.value = value;
}
</script>
<?php
  $id = intval($_COOKIE["SESSIONID"]);
  $res = mysql_query("SELECT `topic` FROM `sessions` WHERE id=".$id);
  $topic = mysql_result($res, 0);
  echo '<a href="#" onclick="populateTopic(\'';
  echo htmlentities($topic, ENT_QUOTES);
  echo '\');">Populate current topic</a>';
?>
\end{lstlisting}

In the PHP code on lines 9-14, the only potentially malicious input value use is on lines 10-13, where a string value from the database (the \texttt{topic} column in the \texttt{sessions} table) is inserted into a JavaScript string inside an \texttt{onclick} handler. Now assume that a malicious user, who can write into the \texttt{topic} column in the \texttt{sessions} table, looks for a way to insert harmful code into the JavaScript content through this input value.

Essential to inserting harmful code is the ability to escape from the context of the string where the input value resides.
Here, this requires inserting the terminating quote character (other contexts may require other characters such as
\texttt{<} or \texttt{\&}, we call these the context switching characters for short).
However, both single quote and double quote are replaced with their respective HTML entities,
\texttt{\&\#39;} or \texttt{\&quot;}, by the \texttt{htmlentities} sanitizer on line 13.
Therefore, the code snippet seems to be safe at the first glance.

The actual vulnerability rests with the fact that the whole \texttt{onclick} attribute value will first be HTML-entity decoded by the client's browser and only then passed to the JavaScript engine. The attacker can insert for example this value into the \texttt{topic} column in the \texttt{sessions} table:

\begin{lstlisting}[numbers=none]
'+alert("Hacked")+'
\end{lstlisting}

The value will be encoded by the \texttt{htmlentities} sanitizer and appear in this form in the HTTP response body:

\begin{lstlisting}[numbers=none]
&#39;+alert(&quot;Hacked&quot;)+&#39;
\end{lstlisting}

The client's browser will decode the HTML entities from the double-quoted HTML attribute value into the original characters before passing the result to the JavaScript engine, which will therefore receive the original input:

\begin{lstlisting}[numbers=none]
'+alert("Hacked")+'
\end{lstlisting}

Even though the terminating quote character was not visibly present in the HTTP response body, the input still successfully breaks out of the single-quoted JavaScript string, and the attacker can execute arbitrary JavaScript code in the client's browser as soon as the \texttt{Populate current topic} link is clicked. This example demonstrates a (stored) context-sensitive XSS flaw, where the malicious XSS payload is incorrectly sanitized by the web application. The \texttt{htmlentities} sanitizer is sufficient for HTML double-quoted attribute values which do not contain embedded JavaScript, but in this example a double sanitization, first with \texttt{addslashes} and only then with \texttt{htmlentities}, was needed.

\medskip

Detecting a context-sensitive XSS flaw is considerably more complicated than detecting a context-insensitive XSS flaw, which appears when an input value is not sanitized at all. To detect a context-insensitive XSS flaw, it is enough to check whether the injected input value appears unchanged in the web application output. In contrast, to detect a context-sensitive XSS flaw, the analysis must also recognize input values that were changed -- encoded -- in the web application output, and decide whether the encoding is sufficient for the browser context where the value appears. This requires a more sophisticated input tracking mechanism and an exhaustive parsing of the HTML document that constitutes the application output. The issue is particularly relevant to blackbox XSS analyzers, whose insight into the data flow inside the application is necessarily limited.

\medskip

Among characteristics shared by both context-sensitive and stored XSS flaws is the fact that their exploits are generally immune to automatic defense mechanisms employed by modern browsers. Browsers based on Chromium~\footnote{Chromium XSS Auditor: \url{https://www.virtuesecurity.com/blog/understanding-xss-auditor}} (e.g. Google Chrome) and some Microsoft browsers~\footnote{Microsoft Internet Explorer and Microsoft Edge XSS Filter: \url{https://blogs.msdn.microsoft.com/ie/2008/07/02/ie8-security-part-iv-the-xss-filter}} provide an automatic XSS protection that looks for unchanged input parameters in the HTML output and therefore prevents context-insensitive reflected XSS attacks. Firefox provides the same functionality as an add-on~\footnote{NoScript Firefox Plugin: \url{https://addons.mozilla.org/en-US/firefox/addon/noscript}}. Modern web browsers thus only prevent XSS exploits of the class that is already systematically covered by state-of-the-art blackbox XSS analyzers, making detection of context-sensitive or stored (or both) XSS flaws more impactful.

\section{Approach}
\label{sec:approach}

The design of our graybox XSS analyzer starts with the classical mechanism used in the detection of context-insensitive reflected XSS flaws, where the analyzer imitates a web browser. Before the analysis, the analyzer is provided with a list of HTTP requests that exercise the web application under analysis (usually, this list is constructed through automated crawling of the web application or by capturing HTTP traffic generated by another web browser). Each request from this list is then analyzed separately -- functionally sensitive parts of the request are identified and mutated by injecting XSS exploit patterns, the modified variations of the request are submitted to the web application, and the application output is checked for the injected patterns. If any injected pattern appears in the output, an XSS flaw was just detected. Among the functionally sensitive parts of the request are obviously the GET and POST parameter values, but also the cookie values, or some HTTP header values such as \texttt{Referer:}.

\subsection{Database Traffic Interception}
\label{subsec:spoofing}

Our graybox analyzer extends the traditional blackbox XSS analyzer functionality by injecting XSS exploit patterns not only into the HTTP requests, but also into the database responses, which in our architecture conceptually constitute additional application input. Technically, this requires solving two problems -- first, recognizing when and what data the application under analysis fetches from the database, and second, injecting the exploit patterns into the database traffic, which is not controlled directly by the analyzer.

To overcome both obstacles, we have designed a database traffic interception protocol,
executed between the analyzer and the database, that can record the database
activity and inject exploit payload into the database responses.

\begin{itemize}
\item In request recording mode, the database carries out the application operations as usual,
but also reports all operations that fetch string values to the analyzer.
We are only interested in operations that fetch string values, because columns of date,
numeric or boolean types cannot realistically carry an XSS exploit payload.
For each string fetch operation, the analyzer is informed about the name of the table (if any, a fetch can also concern temporary tables),
the name of the column, and the value that was fetched.

\item In response injection mode, the analyzer provides the database with an identification of fetch operations whose result should be replaced by an exploit payload (three optional fields -- table name, column name, value fetched), and the actual payload to use instead of the value fetched. The database examines all string fetch operations, and for those that match the specification, the exploit payload is returned instead of the original value.
Other application operations are carried out as usual.
\end{itemize}

With the database traffic interception functionality in place, the work of a classical blackbox analyzer is extended as follows. For each original HTTP request, the analyzer puts the database in the request recording mode and submits the original HTTP request to the application. In addition to the application response, the analyzer also collects the list of all string fetch operations as reported by the database. The fetch operations are then treated as if they were just another kind of application input -- the analyzer puts the database in the response injection mode and for each string fetch operation observed, submits the original HTTP request again while requesting the database to replace the result of that operation with an exploit payload. The application responses are examined in the same way as with other exploit payload injection operations.

In our running example from Section~\ref{sec:example}, the analyzer would first observe the fetch operation on line 10 in the request recording mode.
In the response injection mode, the database would return the exploit payload as the value of the \texttt{topic} column of the \texttt{sessions} table.

\medskip

Including database inputs in the exploit injection increases the number of requests submitted to the application. In practice, the number of distinct database input values consumed by the application when processing a request can be several orders of magnitude higher than the number of HTTP parameters of that request. Treating each database fetch operation separately would therefore require thousands of additional HTTP requests, exceeding practical analysis time. We mitigate the problem by aggregated response injection, which replaces multiple database values in the same request, for example all values fetched from a specific table and column. However, thus reduced injection granularity can lead to false negatives -- it can reduce the code coverage and is more likely to trigger situations where the injected exploit patterns crash the application under analysis. We evaluate the impact of different granularity levels in Section~\ref{sec:eval}.

Although our prototype uses an SQL database, our approach can also handle NoSQL databases -- in principle, the database traffic interception mechanism can treat string fields in a NoSQL exchange same as it treats string columns in an SQL exchange. However, we did not evaluate our approach with a NoSQL database, and suspect that some performance issues in our analysis may be more pronounced with NoSQL databases than with SQL databases. In particular, a potentially more complex database schema would further increase the number of database inputs that need to be intercepted.

\subsection{Exploit Payload Identification}
\label{subsec:payload}

Essential to the analysis is the ability to identify the injected exploit payload in the HTTP responses. Where most blackbox XSS analyzers look only for the unchanged payload, we also want to detect context-sensitive XSS flaws. For that, we need to be able to identify the payload that was inserted into the response body in an encoded form.

Our solution is based on the observation that realistic sanitization routines encode the special characters that can escape browser context and therefore facilitate an exploit (depending on the context that may include \texttt{$<$}, \texttt{"}, \texttt{'}, \texttt{\&}, \texttt{:}, \texttt{\textbackslash} and \texttt{/}), but leave base ASCII alphanumeric characters alone except for perhaps changing the capitalization (these characters are harmless and very frequent, their encoding would therefore needlessly bloat the application output size). Hence we construct an artificial exploit payload from a mixture of special characters and base ASCII alphanumeric characters, and use regular expression matching to identify the base ASCII alphanumeric character sequence used in the payload while skipping over the (possibly encoded) special characters. By relying only on the base ASCII alphanumeric characters, we recognize the exploit payload even after it passes classical encoding algorithms such as HTML entity encoding, URL encoding, JSON escaping, JavaScript literal encoding, CSS value encoding, quote backslashing, and so on.

Even though our artificial exploit payload construction does not produce working XSS exploit patterns,
it suffices for the analysis purposes because it exercises the sanitization functionality
within the application.
The real-life exploit patterns presented throughout our paper are either examples or results of manual analysis.

\medskip

In detail, all our payloads look like \texttt{abcdef$<$gh"ij'kl\&mn:op\textbackslash qr/stuv} -- the letters may differ but the structure and the special characters stay the same (the random alphanumeric prefixes and suffixes reduce the likelihood of erroneous greedy regular expression match). Given a string that constitutes the exploit payload, we then generate the matching regular expression by replacing every base ASCII alphanumeric character by a case insensitive match term (for example the \texttt{(a|A)} term for the \texttt{a} character), and by replacing all other characters with a universal match term \texttt{.*}. To avoid performance issues with very long greedy matches, we limit the match length to 20 characters in the match per one character in the original payload. It is unlikely for the original payload to be encoded so heavily that encoding of any single character would exceed 20 output characters.

As another performance optimization, we use the same regular expression matching mechanism to limit database response injection. We look for the first 20 characters of each database input value in the application output just before that value is intercepted, and only proceed with injecting the exploit payload if the original value is found. If the original database input value does not appear in the output, there is little point in replacing it, which dramatically reduces the number of necessary HTTP requests. The impact of this optimization increases with finer database response injection granularity. The optimization does not help with input strings that are either empty or contain very few base ASCII alphanumeric characters (this should only impact analysis performance but not correctness), and may theoretically introduce false negatives -- however, for this to happen, the application would have to emit the exploit payload although it did not emit the benign original value in the same place, which is unlikely.

\medskip

If we use the \texttt{abcdef$<$gh"ij'kl\&mn:op\textbackslash qr/stuv} exploit payload in our running example from Section~\ref{sec:example},
the code in line 13 will emit a HTML-entity encoded version, embedded in the surrounding HTML content:

\begin{lstlisting}[numbers=none]
<a href="#" onclick="populateTopic('abcdef&lt;gh&quot;
ij&#039;kl&amp;mn:op\qr/stuv');">Populate current topic</a>
\end{lstlisting}

The regular expression derived from the payload, which matches on the HTML content, is:

\begin{lstlisting}[numbers=none]
(a|A)(b|B)(c|C)(d|D)(e|E)(f|F).{0,20}(g|G)(h|H).{0,20}
(i|I)(j|J).{0,20}(k|K)(l|L).{0,20}(m|M)(n|N).{0,20}
(o|O)(p|P).{0,20}(q|Q)(r|R).{0,20}(s|S)(t|T)(u|U)(v|V)
\end{lstlisting}

\subsection{Recursive Output Parsing}

\begin{table*}[!ht]
\centering
\begin{tabular}{|l|l|}
\hline
\makecell[c]{\textbf{Position of the placeholder}} & \makecell[l]{\textbf{Detected sequence of syntax nodes}} \\
\hline
\makecell[l]{\texttt{$<$p style="content:[PH];"$>$}\\\\} & \makecell[l]{[0] HTML double-quoted attribute value, \\ {[1]} CSS property value} \\
\hline
\texttt{$<$b$>$[PH]$<$/b$>$} & [0] HTML text \\
\hline
\texttt{$<$input value=[PH]$>$} & [0] HTML unquoted attribute value \\
\hline
\makecell[l]{\texttt{$<$a href='javascript:call("[PH]");'$>$}\\\\\\} & \makecell[l]{[0] HTML single-quoted attribute value, \\ {[1]} URI with javascript: scheme, \\ {[2]} JavaScript double-quoted string} \\
\hline
\end{tabular}
\caption{Examples of web-browser contexts represented by language-specific syntax node sequences}
\label{tbl:contextexamples}
\end{table*}

After we receive the HTTP response to our mutated request and locate all instances of the injected exploit payload using regular expression matching, we proceed to determine the browser context of each occurrence, required for context sensitive analysis. As the first step, we replace each payload match with a unique placeholder composed of lowercase alphabetical characters. We do this because the exploit payload may have broken the document syntax, which might confuse subsequent parsing, and because we want to avoid automatic decoding of the payload by the invoked parsers.

After the replacement, we essentially imitate a web browser by recursively parsing the HTTP body, and record the context of each placeholder. In a manner similar to~\cite{doi:10.1002/spe.2649}, we represent the context as an ordered sequence of language-specific syntax nodes (nodes that represent the individual syntactic elements of the language grammar) in which the placeholder appeared. Examples of browser contexts are shown in Table~\ref{tbl:contextexamples}.

The process begins with running the HTML output through an HTML parser and looking for HTML syntax nodes containing the placeholders, each recorded browser context therefore begins with an HTML syntax node. If the placeholder appears inside an HTML syntax node that contains embedded content, we decode and extract that embedded content, parse it with the appropriate parser depending on the content language, and determine the encapsulated syntax node inside that embedded language. The placeholder can again reside in an embedded content, in which case we proceed recursively -- see the last example in Table~\ref{tbl:contextexamples}, where HTML embeds URI which embeds JavaScript. Our parsers process all widely supported languages and their nesting with one exception -- we do not attempt to recognize whether a JavaScript string represents an HTML code snippet, URI or its part, CSS code, another JavaScript code (for example in an \texttt{eval} statement) or none of these. Instead, we simply consider all JavaScript strings to be the leaf syntax nodes that do not need further parsing. In reality, a JavaScript string can embed any other language, but detecting such embedded languages is generally an undecidable problem. This makes our context model incomplete, possibly leading to false negatives. As a future work, this functional gap can be mitigated by static or dynamic data-flow analysis of the extracted JavaScript code.

\begin{figure}[!ht]
\centering
\includegraphics{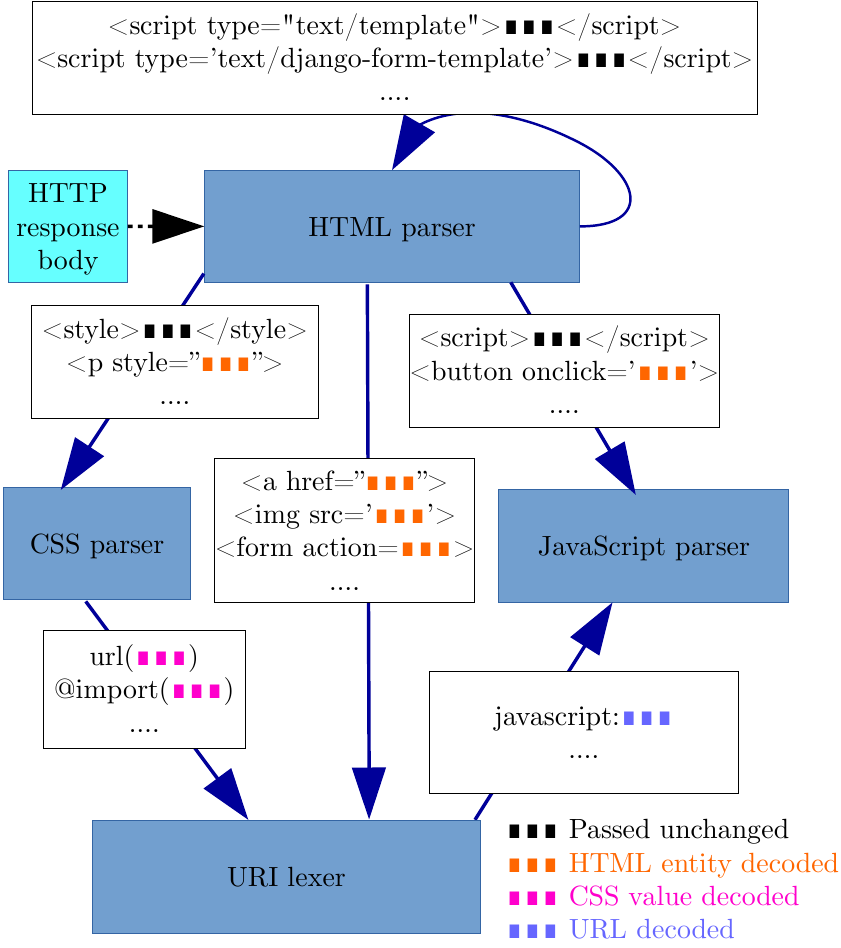}
\caption{Parsers used by our web-browser model and their possible invocations.}
\label{fig:contexts}
\end{figure}

The supported encapsulations of browser languages, parser invocations and decoding algorithms are illustrated in Figure~\ref{fig:contexts}.
In our running example from Section~\ref{sec:example}, the placeholder for the injected payload would be located in a JavaScript context
(the \texttt{onclick} attribute) nested in an HTML context (the response body containing the \texttt{a} element).

\medskip

As an alternative to explicit analysis of the HTTP body, one may consider opening the HTTP response in a real browser, with an extension instrumenting the JavaScript function calls that the exploit payload executes. This fails for multiple reasons, starting with the fact that this would require a universal injection pattern (or a small set of such patterns) that can exploit any context-sensitive XSS flaw. There is no such pattern -- for example, the exploit described in Section~\ref{sec:example} is useful for that particular context mismatch, but it would be useless if the application were to use (also incorrectly) only the \texttt{addslashes} sanitizer. In such case, a working exploit pattern would be:

\begin{lstlisting}[numbers=none]
"><script>alert(String.fromCharCode(72,97,99,107,101,100))</script><
\end{lstlisting}

while the original exploit for the \texttt{htmlentities} sanitizer:

\begin{lstlisting}[numbers=none]
'+alert("Hacked")+'
\end{lstlisting}

would have been made harmless by the \texttt{addslashes} sanitizer alone -- it would simply break the JavaScript syntax by inserting a backslash in front of every single and double quote. In general, every exploitable pair of sanitization and browser context requires a different exploit technique.

As another obstacle to using a real browser, many XSS exploits end up in \texttt{javascript:} URIs or JavaScript \texttt{onevent} handlers. These are not executed immediately after the page load, but require additional user interaction. We need to identify these opportunities automatically, but we cannot automatically test all possible user interactions (at least not unless we resort to model checking with all its implementation and scaling issues). The same goes for constrained attacks such as parameter tampering, where the exploit does not execute a JavaScript function.

Finally, explicit analysis of the HTTP body makes it possible to detect XSS flaws that are exploitable only for some browsers -- for example the \texttt{-moz-binding} injections of JavaScript into CSS\footnote{Moz-binding XSS fun, The Spanner, 2008: \url{http://www.thespanner.co.uk/2008/02/04/moz-binding-xss-fun}} work only with some versions of Firefox and other browsers based on Gecko, other browsers are immune. %

\subsection{Payload Encoding Verification}
\label{sec:levels}

At this stage of the analysis process, we have located all instances of the encoded exploit payload in the web application output and recorded their browser context. Furthermore, we have made sure that the exploit payload contains all typical browser context switching characters -- \texttt{$<$}, \texttt{"}, \texttt{'}, \texttt{\&}, \texttt{:}, \texttt{\textbackslash} and \texttt{/} -- glued together with randomly generated alphabetical strings.

\begin{table*}[!ht]
\centering
\begin{tabular}{|l|l|}
\hline
\textbf{Syntax node} & \textbf{Escape condition} \\
\hline
\makecell[l]{HTML text} & \makecell[l]{Contains \texttt{$<$} character} \\
\hline
\makecell[l]{Double-quoted HTML attribute value} & \makecell[l]{Contains \texttt{"} character} \\
\hline
\makecell[l]{Single-quoted HTML attribute value} & \makecell[l]{Contains \texttt{'} character} \\
\hline
\makecell[l]{HTML data} & \makecell[l]{Contains \texttt{$<$} character and also / \\ character not preceded by a \\ backslash} \\
\hline
\makecell[l]{URI} & \makecell[l]{Contains \texttt{:} character if the \\ placeholder was at the beginning \\ of the URL or \\ \texttt{\&} character if it was elsewhere} \\
\hline
\makecell[l]{Double-quoted JavaScript string literal} & \makecell[l]{After removing all JavaScript literal \\ escape sequences contains \texttt{"} \\ character or \texttt{\textbackslash} character} \\
\hline
\makecell[l]{Single-quoted JavaScript string literal} & \makecell[l]{After removing all JavaScript literal \\ escape sequences contains \texttt{'} \\ character or \texttt{\textbackslash} character} \\
\hline
\makecell[l]{Double-quoted CSS string} & \makecell[l]{After removing all CSS escape \\ sequences contains \texttt{"} character or \texttt{\textbackslash} \\ character} \\
\hline
\makecell[l]{Single-quoted CSS string} & \makecell[l]{After removing all CSS escape \\ sequences contains \texttt{'} character or \texttt{\textbackslash} \\ character} \\
\hline
\makecell[l]{All other syntax nodes} & \makecell[l]{True} \\
\hline
\end{tabular}
\caption{Escape conditions indicating security flaw in given language-specific syntax node.}
\label{tbl:whitelist}
\end{table*}

\begin{table*}[!ht]
\centering
\begin{tabular}{|l|l|}
\hline
\textbf{Syntax node} & \textbf{Decoding algorithm} \\
\hline
\makecell[l]{HTML text} & None (always leaf syntax node) \\
\hline
\makecell[l]{Double-quoted HTML attribute value} & HTML entity decoding \\
\hline
\makecell[l]{Single-quoted HTML attribute value} & HTML entity decoding \\
\hline
\makecell[l]{HTML data} & None (kept identical) \\
\hline
\makecell[l]{URI} & URL decoding \\
\hline
\makecell[l]{Double-quoted JavaScript string literal} & None (always leaf syntax node) \\
\hline
\makecell[l]{Single-quoted JavaScript string literal} & None (always leaf syntax node) \\
\hline
\makecell[l]{Double-quoted CSS string} & CSS-value decoding \\
\hline
\makecell[l]{Single-quoted CSS string} & CSS-value decoding \\
\hline
\makecell[l]{All other syntax nodes} & None (always leaf syntax node) \\
\hline
\end{tabular}
\caption{Decoding applied to content in language-specific syntax node.}
\label{tbl:decodings}
\end{table*}

To determine whether the exploit payload was sufficiently sanitized, we simulate the individual browser decoding steps on the payload from the outermost to the innermost syntax node (thus beginning with an HTML syntax node). At each step, we consult a list of language-specific escape conditions in Table~\ref{tbl:whitelist} for that syntax node and report an XSS flaw if the escaping condition is true. If not, we apply a language-specific decoding algorithm from the list in Table~\ref{tbl:decodings} and proceed to the next syntax node. If the syntax node is not on our list (such as an HTML tag name or a JavaScript variable name), we immediately report a potential XSS flaw and resort to manual analysis. This approach is in line with OWASP recommendations~\footnote{XSS (Cross Site Scripting) Prevention Cheat Sheet, OWASP: \\ \url{https://www.owasp.org/index.php/XSS\_Prevention\_Cheat\_Sheet}} and with the fact that sanitization in non-standard browser contexts is often error-prone.

We further refine the classification by looking at how the discovered bugs can be exploited.
We report an XSS flaw that \emph{does not permit JavaScript execution} for a payload that escapes into a URL value but not at its beginning
-- such flaws permit only parameter tampering~\footnote{Parameter Tampering: \url{https://www.owasp.org/index.php/Web\_Parameter\_Tampering}}.
Based on our experience and on other sources~\cite{Weinberger:2011:SAX:2041225.2041237}, this XSS flaw pattern is the most frequent one in existing applications.
We report an XSS flaw that \emph{possibly permits arbitrary JavaScript execution} when a JavaScript string contains only an unescaped backslash character
-- such flaws can always break syntax and thus cause denial of service, but are usually not exploitable towards arbitrary JavaScript execution.
All other flaws are reported as flaws that \emph{permit arbitrary JavaScript execution}
-- in our experience such flaws can either be exploited to execute arbitrary JavaScript
or occur very rarely and are thus worthy of manual analysis.

\medskip

\begin{figure}[!ht]
\centering
\includegraphics{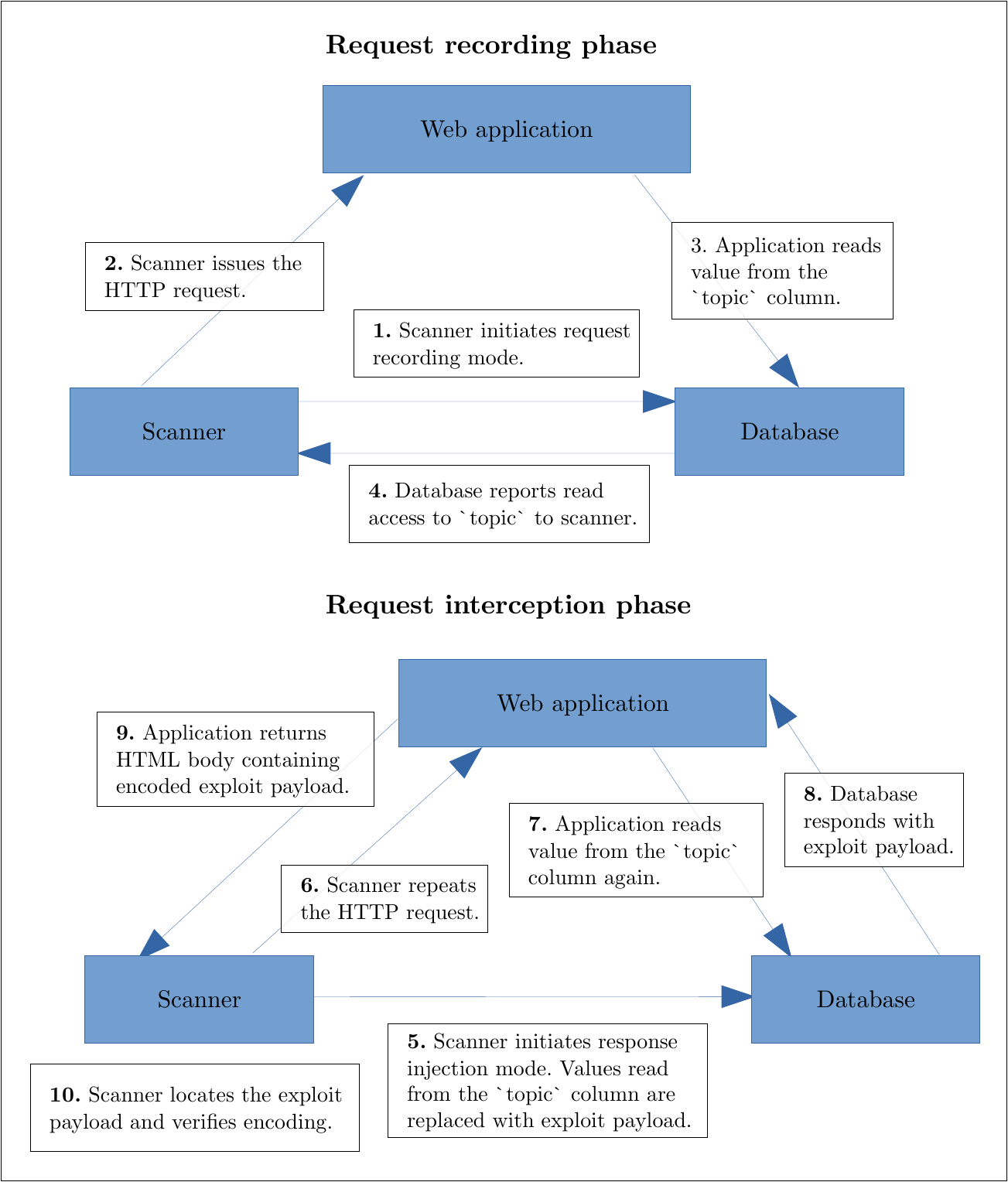}
\caption{XSS flaw detection algorithm applied to the running example from Section~\ref{sec:example}.}
\label{fig:flowingexample}
\end{figure}

Returning for the last time to our running example from Section~\ref{sec:example},
recall that the exploit payload was located in a single-quoted JavaScript literal
nested in a double-quoted HTML attribute value:

\begin{lstlisting}[numbers=none]
<a href="#" onclick="populateTopic('abcdef&lt;gh&quot;
ij&#039;kl&amp;mn:op\qr/stuv');">Populate current topic</a>
\end{lstlisting}

We therefore first check whether the payload contains the \texttt{"} character,
whose presence would indicate the attacker is able to prematurely terminate
the attribute value string and escape the HTML attribute value context.
It does not, we therefore decode the value as HTML entity:

\begin{lstlisting}[numbers=none]
populateTopic('abcdef<gh"ij'kl&mn:op\qr/stuv');
\end{lstlisting}

We next check whether the decoded payload contains the \texttt{'} or \texttt{\textbackslash} characters.
It does, both, hence we report an XSS flaw that permits arbitrary JavaScript execution.
An end-to-end overview of the steps is displayed
in Figure~\ref{fig:flowingexample}.

\subsection{Prototype Implementation}

We have implemented our prototype graybox analyzer as a plugin for the OWASP Zed Attack Proxy (ZAP).\footnote{The OWASP Zed Attack Proxy: \url{https://www.zaproxy.org}} For HTML parsing, we have extended the Jsoup parser~\footnote{jsoup: Java HTML Parser: \url{https://jsoup.org}} to distinguish between single-quoted, double-quoted and unquoted HTML attribute values. For the parsing of CSS sheets and declarations, we have similarly extended the SteadyState CSS parser~\footnote{CSS Parser: \url{http://cssparser.sourceforge.net}} to differentiate between single-quoted, double-quoted and unquoted CSS strings and URI values. We have used Rhino~\footnote{Rhino: \url{https://developer.mozilla.org/en-US/docs/Mozilla/Projects/Rhino}} for parsing Javascript, here the distinction between single-quoted and double-quoted string literals is already supported. Finally, we have implemented the URI lexer from scratch.
We note that the exact choice of parsers may affect the analysis results especially where subtle differences in treating potentially
malformed content are concerned, however, our results show that the prototype implementation suffices to evaluate the approach.

As the general purpose database connected to the analyzer we have extended MariaDB,\footnote{MariaDB.org: Supporting Continuity and Open Collaboration: \url{https://mariadb.org}} a binary replacement to MySQL. We instrument the string fetch operations in the MariaDB network protocol handler,\footnote{COM\_QUERY command, MariaDB: \url{https://mariadb.com/kb/en/library/com\_query/}} that is, the insertion of \texttt{MYSQL\_TYPE\_VARCHAR}, \texttt{MYSQL\_TYPE\_VAR\_STRING} and \texttt{MYSQL\_TYPE\_STRING} values into the \texttt{ResultSet}\footnote{Server response packet ResultSet, MariaDB: \url{https://mariadb.com/kb/en/library/resultset/}} server packet. Therefore, our prototype is compatible with every web application that supports the MySQL client-server protocol\footnote{MySQL Client/Server Protocol, MySQL: \url{https://dev.mysql.com/doc/internals/en/client-server-protocol.html}} as a client. MySQL (together with MariaDB) is currently probably the most popular DB architecture~\footnote{Most popular databases in 2018 according to StackOverflow survey: \\ \url{https://www.eversql.com/most-popular-databases-in-2018-according-to-stackoverflow-survey/}}.

For readers interested in technical details of our prototype implementation, we provide complete source code of all the patches made to ZAP, MariaDB, Jsoup and the SteadyState CSS parser\footnote{\url{https://bit.ly/2V16dN7}}~\footnote{\url{https://bit.ly/2USLAm6}}~\footnote{\url{https://bit.ly/2E9FiZH}}~\footnote{\url{https://bit.ly/2S31nNm}}~\footnote{\url{https://bit.ly/2BA4OFU}}.

\section{Evaluation}
\label{sec:eval}

To evaluate our approach in sufficiently challenging conditions, we have decided to aim for more mature applications. While this may bias our selection towards older coding practices, we want to avoid applications that only had short time to be scrutinized by security tools and security experts. Specifically, we have considered applications that are 1) open-source, 2) developed for at least five years, 3) actively maintained, 4) designed to handle security-sensitive data and 5) have at least one thousand publicly registered users of their code. Section~\ref{subsec:existing} shows that even these precautions were not entirely sufficient to avoid flaws automatically detectable by state-of-the-art tools, however, at least half of the selected applications did appear flaw-free.
We have also considered making apples-to-apples comparison on some of the applications used for evaluation by authors of related work in Section ~\ref{sec:relwork}, however, these use much older code than our selected applications. Web application technologies change very rapidly and an evaluation using applications that, by now, may not even run, has little practical value. For similar reasons, we have avoided synthetic benchmarks with artificially created security flaws.

From the applications matching all five criteria, we picked eight applications in order to cover various security-relevant use cases and various technologies. Our picks are (A1) Joomla,\footnote{Joomla Content Management System: \url{https://www.joomla.org}} a general-purpose CMS written in PHP, (A2) OrchardCMS,\footnote{Orchard CMS: \url{http://orchardproject.net}} formerly called Microsoft Oxite, a blog engine written in ASP.NET MVC, (A3) SuiteCRM,\footnote{SuiteCRM, Open Source CRM for The World: \url{https://suitecrm.com}} an open-source fork of SugarCRM, a bussiness-data management CRM written in PHP, (A4) Fat Free CRM,\footnote{Fat Free CRM: Ruby On Rails-Based Open Source CRM Platform: \url{http://www.fatfreecrm.com}} a general-purpose CRM written in Ruby on Rails, (A5) OpenEMR,\footnote{OpenEMR, The World's Leading Open-Source Electronic Medical Record and Practice Management Software: \url{https://www.open-emr.org}} an ONC Complete Ambulatory EHR certified health record management CMS written in PHP, (A6) Jeesite,\footnote{Jeesite Rapid Development Platform: \url{http://jeesite.com}} a software development CMS written in JSP and Java, (A7) PrestaShop,\footnote{Create and Develop Your Business with PrestaShop: \url{https://www.prestashop.com}} an on-line shop CMS written in PHP and (A8) Mezannine,\footnote{Mezzanine - The Best Django CMS: \url{http://mezzanine.jupo.org}} a general purpose CMS written in Django and Python.

We have installed each application and populated it with sample or demo data using an autopopulation mechanism that was provided by the application, or we extracted sample data from a public demo instance of that application. If there were multiple autopopulation options, we always picked the most complex autopopulation scenario or the one dedicated to application testing. Afterwards, we have used the ZAP spider with its default settings (sending both GET and POST forms, recursive traversal depth limit 5, no limit on the number of children, accept cookies, send referrers, consider both parameter names and values) for crawling the application with HTTP requests (we did not use the AJAX spider extensions, limiting our analysis to standard requests). We provided the spider with log-in form requests and log-out criteria, so that it could also access links that require authentication. Finally, we tweaked the web application and spider settings in order to optimize the spider behavior for the particular application under test:

\begin{itemize}
\item In SuiteCRM we made the \texttt{config.php} file non-writable by the Apache user and blacklisted the backup handling URL paths, because otherwise the spider restored an artificially generated invalid backup of the database and destroyed the application.
\item In Mezzanine we manually disabled the CSRF protection middleware in Django settings, because the spider was not able to handle authentications protected by that middleware. 
\item In Jeesite we blacklisted the URL paths for user administration, because otherwise the spider would change the password of the user under which it was logged in to a random value and stay logged off for the rest of the crawl.
\item In OpenEMR we manually rewrote the generated \texttt{.htaccess} files in order to enable directory listing, because otherwise the spider was unable to discover a significant amount of original links.
\item For OrchardCMS we completely disabled caching of ORM objects in order to force the application to always fetch fresh values from the database.
\end{itemize}

After the spider crawled the application, we have backed up the generated ZAP context including the list of original HTTP requests, the MariaDB database state of the application and the application installation directory. We have restored the complete backup before each scan to prevent interference between individual scans. An active scan with our plugin was started afterwards.

We have executed all the analyzed applications on a Fedora Linux server with an Intel Xeon E5345 CPU (4 cores, 8 threads, base clock 2.33 GHz) and 8 GB RAM. PHP applications were executed on Apache 2.4.25 with PHP version 7.0.25, Fat Free CRM was executed on Rails 5.1.6 with Ruby 2.4.4p296, Jeesite was executed on Apache Tomcat 7.1.10 with OpenJDK 1.8.0\_151 and Mezannine on Python 2.7.13 and Django 1.11.15. For OrchardCMS we used a special deployment configuration. We have executed the application inside a VirtualBox instance with Windows 10 Enterprise (1 core and 2GB RAM), in which we have installed OrchardCMS using Microsoft Web Deploy 4.0. Further, we have configured it to connect over TCP to the MariaDB database running on the Linux host machine together with ZAP and our analyzer plugin. For OrchardCMS, we also needed to disable ORM caching to avoid reusing database fetch results by the .NET runtime. Therefore, any interpretation of the performance metrics from the OrchardCMS analysis should take into account this atypical deployment scenario.

\subsection{Analysis Results}

\begin{table*}[!ht]
\centering
\begin{tabular}{|l|r|r|r|r|r|r|r|r|}
\hline
& \multicolumn{1}{c|}{A1} & \multicolumn{1}{c|}{\makecell[c]{A2}} & \multicolumn{1}{c|}{\makecell[c]{A3}} & \multicolumn{1}{c|}{\makecell[c]{A4}} & \multicolumn{1}{c|}{\makecell[c]{A5}} & \multicolumn{1}{c|}{A6} & \multicolumn{1}{c|}{\makecell[c]{A7}} & \multicolumn{1}{c|}{\makecell[c]{A8}} \\
\hline
Programming language & \makecell[c]{PHP} & \makecell[c]{C\# \\ Razor} & \makecell[c]{PHP} & \makecell[c]{Ruby \\ HAML} & \makecell[c]{PHP} & \makecell[c]{Java \\ JSP} & \makecell[c]{PHP} & \makecell[c]{Python \\ Django} \\
\hline
Version & Oct 30 & Nov 14 & Oct 20 & Oct 26 & Oct 28 & Oct 18 & Nov 8 & Nov 11 \\
\hline
Lines of code & 755k & 249k & 1050k & 36k & 794k & 47k & 406k & 21k \\
\hline
Correct sanitizations & 97 & 117 & 588 & 84 & 421 & 33 & 442 & 29 \\
\hline
Incorrect sanitizations & 22 & 3 & 31 & 20 & 83 & 11 & 29 & 8 \\
\hline
\makecell[l]{Incorrectly classified \\ correct sanitizations (FP)} & 0 & 0 & 0 & 0 & 0 & 0 & 0 & 0 \\
\hline
\end{tabular}
\caption{Correct and incorrect sanitization instances summary. Version dates denote commits on master branch and are all from 2018.}
\label{tbl:correct}
\end{table*}

The overall analysis results are given in~\autoref{tbl:correct}. As a very basic observation, we can see that the analyzer never misclassified a correct sanitization for an incorrect one, and identified a share of incorrect sanitizations in every application. However, for an accurate understanding of the results, we need to delve further into the classification of incorrect sanitizations. Because our analysis is motivated by possible security exploits, it is essential to also determine whether a reported incorrect sanitization can be exploited and how. We do this in~\autoref{tbl:exploits}, where the severity levels from Section~\ref{sec:levels} are combined with a manual post-hoc analysis that distinguishes three situations -- one when we have been able to exploit the identified flaw using only the web application user interface, one when we have been able to exploit the flaw but only with direct database access, and one when we did not manage to exploit the flaw, or where the flaw was clearly not exploitable. Note that since manual analysis is involved (we only count flaws which we could exploit ourselves), the numbers for manually exploited flaws are a conservative lower bound.

\begin{table*}[!ht]
\centering
\begin{tabular}{|l|r|r|r|r|r|r|r|r|}
\hline
& \makecell[c]{A1} & \makecell[c]{A2} & \makecell[c]{A3} & \makecell[c]{A4} & \makecell[c]{A5} & \makecell[c]{A6} & \makecell[c]{A7} & \makecell[c]{A8} \\
\hline
\multicolumn{9}{l}{\textbf{Incorrect sanitizations reported as permitting JavaScript execution}} \\
\hline
\makecell[l]{manually exploited for arbitrary \\ JavaScript execution} & 5 & 1 & 3 & 1 & 37 & 5 & 12 & 2 \\
\hline
\makecell[l]{manually exploited for arbitrary JavaScript \\ execution only when writing directly to \\ the database} & 6 & 0 & 2 & 0 & 4 & 1 & 0 & 0 \\
\hline
\makecell[l]{manual exploit for arbitrary JavaScript \\ execution not found} & 0 & 0 & 0 & 0 & 0 & 0 & 0 & 0 \\
\hline
\multicolumn{9}{l}{\textbf{Incorrect sanitizations reported as possibly permitting JavaScript execution}} \\
\hline
\makecell[l]{manually exploited for arbitrary \\ JavaScript execution} & 0 & 1 & 1 & 0 & 2 & 0 & 1 & 0 \\
\hline
\makecell[l]{manually exploited for arbitrary JavaScript \\ execution only when writing directly to \\ the database} & 0 & 0 & 1 & 0 & 1 & 0 & 0 & 0 \\
\hline
\makecell[l]{manual exploit for arbitrary JavaScript \\ execution not found} & 0 & 0 & 5 & 0 & 8 & 0 & 3 & 0 \\
\hline
\multicolumn{9}{l}{\textbf{Incorrect sanitizations reported as not permitting JavaScript execution}} \\
\hline
\makecell[l]{manually exploited for arbitrary \\ JavaScript execution} & 0 & 0 & 0 & 0 & 0 & 0 & 0 & 0 \\
\hline
\makecell[l]{manually exploited for arbitrary JavaScript \\ execution only when writing directly to \\ the database} & 0 & 0 & 0 & 0 & 0 & 0 & 0 & 0 \\
\hline
\makecell[l]{no exploit for arbitrary JavaScript \\ execution possible} & 11 & 1 & 19 & 19 & 31 & 5 & 13 & 6 \\
\hline
\end{tabular}
\caption{Automatically assigned severity vs manually investigated exploitability of incorrect sanitizations.}
\label{tbl:exploits}
\end{table*}

\medskip

To summarize \autoref{tbl:exploits}, the 207 cases of incorrect sanitization include 71 cases (34\%) that were confirmed as permitting arbitrary JavaScript execution through the user interface and 15 cases (7\%) that permit such execution only with direct database access. There were 105 cases (51\%) that were confirmed as not permitting arbitrary JavaScript execution. In 16 cases (8\%) we did not find a manual exploit despite the automated analysis reporting one may exist.

We can also view the numbers in~\autoref{tbl:exploits} from a strictly practical security perspective, looking only at the number of reports that have to be investigated. Our tool flagged 102 incorrect sanitization instances as worthy of attention, with 79 classified as exploitable and 23 classified as possibly exploitable. Following this report, a security expert would find that at least 70\% of all the flagged flaws were XSS flaws exploitable for arbitrary JavaScript execution, and the tool flagged 93\% of those as such.

In our experience, the manual analysis of incorrect sanitization instances was mostly straightforward. In the case of reflected XSS flaws, the analyst must manually verify those incorrectly sanitized values that end up in JavaScript strings and check whether there is a dataflow path between the string and the HTML output (unencoded backslashes can be used to encode an arbitrary HTML code using the JavaScript \texttt{\textbackslash xHH} encoding into the JavaScript string content). For stored XSS flaws, the analyst receives a report that contains the URL, the source table name and the column name (for example table \textit{articles}, column \textit{title}) the payload emitted on the output, and the browser context of the payload (displayed as a comparison of unencoded and encoded string containing the escape characters). Given this data, the analyst can typically easily check whether the required escape character can be inserted into the relevant database column, because most values can only be changed in a limited number of ways through the user interface (for example changing the title of the article in the database column). In our evaluation, we only had to consider multiple insertion methods in one case in SuiteCRM and in two cases in OpenEMR. Automating this part of the analysis is, in our opinion, not feasible, because human understanding of the logical application structure is crucial for the analysis.

\medskip

For the interested reader we provide videos showing exploits of all 71 discovered XSS flaws that we exploited with an arbitrary JavaScript execution.\footnote{\url{https://drive.google.com/open?id=1Wrr-OOxI7BE7cuFqjwQXAui3glFuvVrS}} As a part of responsible disclosure, we have reported all these exploits to the maintainers of the affected applications, all confirmed their acceptance more than 90 days ago. Some of the maintainers have applied for and received CVE records for the flaws that we discovered in their applications (CVE-2018-1000842, CVE-2019-6261, CVE-2019-6264, CVE-2019-7741 or CVE-2019-7744). We also reported the discovered XSS flaws to the maintainers of popular forks that inherited the vulnerable code (for example, we reported the XSS flaws discovered in PrestaShop also to Thirty Bees~\footnote{Thirty Bees Open Source eCommerce Platform - eCommerce That Works: https://thirtybees.com}, because the Thirty Bees codebase was vulnerable as well). At the time of this writing, all the flaws we have identified in Joomla, OrchardCMS, Fat Free CRM, Jeesite and Mezzanine, and some of the flaws we have identified in OpenEMR and PrestaShop, were resolved. 

\medskip

To compare our prototype with the existing tools and to evaluate the overall usefulness, we will further focus strictly on those (71) incorrect sanitization instances that are exploitable and allow an arbitrary JavaScript execution without direct database access.

\subsection{Discovered XSS Flaw Patterns}

\begin{table*}[!ht]
\centering
\begin{tabular}{|l|r|r|r|r|r|r|r|r|}
\hline
& \multicolumn{1}{c|}{A1} & \multicolumn{1}{c|}{\makecell[c]{A2}} & \multicolumn{1}{c|}{\makecell[c]{A3}} & \multicolumn{1}{c|}{\makecell[c]{A4}} & \multicolumn{1}{c|}{\makecell[c]{A5}} & \multicolumn{1}{c|}{A6} & \multicolumn{1}{c|}{\makecell[c]{A7}} & \multicolumn{1}{c|}{\makecell[c]{A8}} \\
\hline
Reflected context-insensitive & 0 & 0 & 0 & 0 & 5 & 0 & 7 & 0 \\
Reflected context-sensitive & 0 & 0 & 2 & 0 & 6 & 4 & 3 & 0 \\
Stored context-insensitive & 0 & 1 & 1 & 0 & 9 & 0 & 1 & 0 \\
Stored context-sensitive & 5 & 1 & 1 & 1 & 19 & 1 & 2 & 2 \\
\hline
\end{tabular}
\caption{Bug patterns of incorrect sanitizations exploitable with an arbitrary JavaScript execution.}
\label{tbl:exploitable}
\end{table*}

Table~\ref{tbl:exploitable} classifies the discovered XSS flaw patterns. The results confirm our assumption that the detection of stored and context-sensitive XSS flaws is more impactful than the detection of reflected and context-insensitive XSS flaws. While only two analyzed applications (OpenEMR and PrestaShop) contained classical reflected context-insensitive XSS flaws that are covered by state-of-the-art blackbox XSS scanners and by modern browser XSS prevention mechanisms, all eight analyzed applications contained stored context-sensitive XSS flaws. Moreover, three of the analyzed applications (Joomla, Fat Free CRM and Mezzanine CMS) contained only XSS flaws that are both stored and context-sensitive at the same time.

\subsection{Analysis Performance and Response Injection Granularity}

\begin{table*}[!ht]
\centering
\begin{tabular}{|l|r|r|r|r|r|r|r|r|}
\hline
& \multicolumn{1}{c|}{A1} & \multicolumn{1}{c|}{\makecell[c]{A2}} & \multicolumn{1}{c|}{\makecell[c]{A3}} & \multicolumn{1}{c|}{\makecell[c]{A4}} & \multicolumn{1}{c|}{\makecell[c]{A5}} & \multicolumn{1}{c|}{A6} & \multicolumn{1}{c|}{\makecell[c]{A7}} & \multicolumn{1}{c|}{\makecell[c]{A8}} \\
\hline
\rowcolor[gray]{.9}
\multicolumn{9}{|c|}{Default ZAP plugins for reflected and stored XSS detection, v33.0} \\
\hline
\rowcolor[gray]{.9} HTTP requests & 583092 & 3826 & 95539 & 46818 & 240854 & 18225 & 49958 & 2532 \\
\rowcolor[gray]{.9} Analysis time & 13h56m & 1h10m & 4h49m & 6h8m & 11h46m & 1h22m & 3h44m & 10m \\
\rowcolor[gray]{.9} XSS discoveries & 0 & 0 & 0 & 0 & 7 & 0 & 8 & 0 \\
\hline
\multicolumn{9}{|c|}{Injecting individual fetches (specific table and column and value)} \\
\hline
HTTP requests & 179\% & 93\% & 148\% & 80\% & 173\% & 147\% & 98\% & 120\% \\
Analysis time & 699\% & 361\% & 572\% & 312\% & 677\% & 499\% & 383\% & 550\% \\
XSS discoveries & 5 & 2 & 4 & 1 & 39 & 5 & 13 & 2 \\
\hline
\multicolumn{9}{|c|}{Grouping all fetches from the same table and column (specific table and column)} \\
\hline
HTTP requests & 66\% & 57\% & 62\% & 53 \% & 68 \% & 55\% & 57\% & 57\% \\
Analysis time & 254\% & 226\% & 245\% & 212 \% & 268 \% & 188\% & 220\% & 280\% \\
XSS discoveries & 4 & 2 & 4 & 1 & 37 & 5 & 13 & 2 \\
\hline
\multicolumn{9}{|c|}{Grouping all fetches from the same table (specific table)} \\
\hline
HTTP requests & 53\% & 55\% & 56\% & 52\% & 54\% & 53\% & 52\% & 52\% \\
Analysis time & 208\% & 217\% & 216\% & 207\% & 211\% & 172\% & 204\% & 250\% \\
XSS discoveries & 3 & 1 & 1 & 0 & 32 & 5 & 13 & 2 \\
\hline
\multicolumn{9}{|c|}{Injecting all database fetches together} \\
\hline
HTTP requests & 51\% & 54\% & 55\% & 52\% & 51\% & 51\% & 51\% & 51\% \\
Analysis time & 201\% & 210\% & 211\% & 205\% & 198\% & 162\% & 198\% & 230\% \\
XSS discoveries & 0 & 1 & 1 & 0 & 22 & 4 & 11 & 1 \\
\hline
\end{tabular}
\caption{Performance statistics on different levels of response injection granularity. All XSS discovery counts are absolute values, performance related values are absolute for baseline and relative to baseline otherwise.}
\label{tbl:performance}
\end{table*}

Table~\ref{tbl:performance} gives the basic performance metrics of our evaluation analysis when carried out with different database response injection granularities, contrasted against the baseline of the default ZAP plugins for reflected and stored XSS detection (version 33.0 of the Active scan rules in the ZAP extensions project\footnote{OWASP ZAP Addons: \url{https://github.com/zaproxy/zap-extensions}}) executed with parallelism level set to 5. Our choice of evaluated granularity levels was motivated by ease of implementation -- while we can come up with other combinations, we see no reason to believe any particular combination should provide benefits broadly applicable across multiple applications.

Our analysis is always slower than the baseline but never prohibitively so. We attribute this to the fact that our solution is single-threaded by design. If we want to retain the nearly blackbox character of our analysis, the state of the database during request recording and response injection must be considered global as we cannot differentiate between application request processing threads and match them to the database requests. For certain frameworks (such as uWSGI), it should be possible to track the arriving HTTP requests all the way to the database, but that would make our analysis platform dependent and again violate the nearly blackbox character. The default ZAP plugins of the baseline allow conceptually almost unlimited parallelism and therefore achieve lower average time per request.

Table~\ref{tbl:performance} also shows that the best recall was achieved with the finest granularity response injection. The best ratio between the number of discovered XSS exploits and the analysis time was achieved when all database fetch operations from the same table and column were intercepted together. This performance advantage may increase with naturally populated applications, whose database tables may contain many more rows than our autopopulated installations. However, even such small coarsening of response injection granularity introduces missed XSS exploits.

\subsection{Comparison with Existing XSS Scanners}
\label{subsec:existing}

\begin{table*}[!ht]
\centering
\begin{tabular}{|l|r|r|r|r|r|r|r|r|}
\hline
& \multicolumn{1}{c|}{A1} & \multicolumn{1}{c|}{\makecell[c]{A2}} & \multicolumn{1}{c|}{\makecell[c]{A3}} & \multicolumn{1}{c|}{\makecell[c]{A4}} & \multicolumn{1}{c|}{\makecell[c]{A5}} & \multicolumn{1}{c|}{A6} & \multicolumn{1}{c|}{\makecell[c]{A7}} & \multicolumn{1}{c|}{\makecell[c]{A8}} \\
\hline
Acunetix Web Vulnerability Scanner & 0 & 0 & 0 & 0 & 10 & 0 & 8 & 0 \\
Burp Suite Professional & 0 & 0 & 2 & 0 & 19 & 3 & 10 & 0 \\
OWASP ZAP XSS plugins & 0 & 0 & 0 & 0 & 7 & 0 & 8 & 0 \\
Our ZAP plugin & 5 & 2 & 4 & 1 & 39 & 5 & 13 & 2 \\
\hline
\end{tabular}
\caption{Comparison of XSS flaw counts as discovered by our analysis and three state-of-the-art blackbox XSS scanners.}
\label{tbl:comparison}
\end{table*}

For the comparative evaluation of our approach, we have again used the default OWASP ZAP extensions, along with two commercial products classified in an extensive blackbox web vulnerability comparison~\cite{10.1007/978-3-642-14215-4_7} among the highest ranking and relatively mutually orthogonal. Those are Burp Suite\footnote{Burp Suite Scanner - PortSwigger: \url{https://portswigger.net/burp}} and Acunetix.\footnote{Website Security - Keep in Check with Acunetix: \url{http://www.acunetix.com}}. We have used the Active scan rules version 33.0 from OWASP ZAP extensions and Burp Suite Professional 1.7.32.

We have provided Acunetix with the log-in sequences, Burp Suite with session definitions and both applications with sitemaps generated by the ZAP spider. Afterwards, we have started scans aimed at the detection of reflected and stored XSS flaws.

The comparison results are given in Table~\ref{tbl:comparison}. PrestaShop and OpenEMR contained XSS flaws that were discovered by all four scanners - mostly reflected context-insensitive XSS flaws. Our assumption that popular and certified applications that handle security-sensitive data (here health records and purchase records including credit card numbers) should not contain easily discoverable XSS flaws was incorrect in those two cases. Notably, Burp Suite Professional scored on a subset of reflected context-sensitive XSS flaws in SuiteCRM and Jeesite -- these were insertions of HTML entity encoded GET or POST parameter values into \texttt{onevent} handlers. Finally, the four remaining applications (Joomla, OrchardCMS, Fat Free CRM and Mezzanine) could claim state-of-the-art level of security, because OWASP ZAP, Burp Suite and Acunetix did not find any XSS flaws in them. It is worth mentioning that Acunetix has special support for detection of XSS flaws in Joomla and even scans Joomla using a database of fingerprints that identify security bugs discovered in Joomla in the past\footnote{Joomla! Ensures Website Security with Acunetix: \url{https://www.acunetix.com/vulnerability-scanner/cs\_joomla}}~\footnote{Joomla Vulnerability Scanner: Enter Acunetix!: \url{https://www.acunetix.com/vulnerability-scanner/joomla-vulnerability-scanner}}.

Our ability to find XSS flaws in Joomla, OrchardCMS, Fat Free CRM and Mezzanine, which the other scanners all declared free of XSS flaws, indicates that our approach surpasses the state-of-the-art level of blackbox XSS security testing in the ability to detect stored and context-sensitive XSS flaws.

\section{Related Work}
\label{sec:relwork}

We are not aware of any existing tools that shortcut the dataflow analysis of stored XSS flaws by intercepting the communication with the database while avoiding whitebox analysis of the web application. However, existing solutions may share isolated aspects of our approach -- some intercept the database traffic, some avoid whitebox analysis, some detect context-sensitive XSS flaws. In each of the following subsections we focus on one shared aspect.

\subsection{Database Traffic Interception}

Sentinel~\cite{li2012sentinel} is similar to our approach because of its focus on the interface between a web application and its database. It works in two modes. In the learning mode Sentinel observes the typical application execution workflow and extracts a set of invariants from the SQL queries that are generated by the web application. Eventually, Sentinel is switched from the learning mode to the testing mode. Then, when the application generates an SQL query that violates the extracted invariants, the query is identified as a potential attack attempt and rejected. However, we handle the interface between a web application and its database very differently from Sentinel. We do not merely observe the communication between the web application and the database, but also actively modify the traffic. Our approach also works on a lower level of the software stack. We do not analyze the SQL language or any of its dialects, and instead operate directly on the network protocol between the web application and the database. Therefore, we support also applications using ORM or other alternative interfaces instead of SQL. An approach similar to Sentinel is also used by BLOCK~\cite{li2011block}. It also works in a learning and a testing mode. BLOCK operates on a higher level of abstraction than Sentinel, but it is limited to state violation attacks. It finds and enforces invariants over session identifier variables instead of SQL queries.

\subsection{Non-Whitebox Detection of XSS Vulnerabilities}

In~\cite{7412085}, Parvez et al. tested three state-of-the-art blackbox analyzers on a relatively simple stored context-insensitive XSS flaw without any non-trivial obstacles (such as JavaScript, captchas or complex form value consistency checks) and found that even this simple scenario presented a big challenge for the tested blackbox XSS scanners. Benchmarking of eight state-of-the-art blackbox XSS scanners done by Bau et al. in~\cite{5504795} showed that all tested XSS scanners were quite successful when looking for reflected XSS vulnerabilities, but their ability to detect stored XSS vulnerabilities was considerably weaker. An even more systematic benchmarking of eleven state-of-the-art blackbox XSS scanners presented by Doupe et al. in~\cite{10.1007/978-3-642-14215-4_7} led to the same conclusions. All three sources show that our approach exceeds the state-of-the-art level of blackbox XSS testing by removing the gap between reflected XSS flaw and stored XSS flaw detection.

Multiple previous attempts tried to increase the effectiveness of blackbox XSS scanners. Classical web crawlers assume that web applications are stateless. This assumption is obviously unsound from the security perspective. The state-aware blackbox scanner~\cite{doupe2012enemy} presented again by Doupe et al. tries to heuristically infer the state of the web application (e.g. when the user is logged in, the web application is in a different state than when the user is logged out) and cover the web application URLs in all possible web application states. This approach was evaluated on eight web applications, showing better detection capabilities on two of those applications when compared with classical stateless blackbox scanners. Pellegrino and Balzarotti in~\cite{pellegrino2014toward} use a similar approach in order to detect logical security flaws in web applications (examples of logical flaws are a possibility to shop in an online shop without paying or a possibility to change the administrator password without being logged in as an administrator).

\subsection{Non-Whitebox Detection of Context-Sensitive XSS Vulnerabilities}

Non-whitebox detection of context-sensitive XSS flaws is performed by Burp Suite,\footnote{Burp Suite: \url{https://portswigger.net/burp}} which we used in the evaluation section. Burp Suite is a proprietary software which supports HTML decoding and is able to detect some sanitization mismatches like inserting HTML-entity encoded input values into JavaScript \texttt{onevent} handlers.\footnote{XSS: Beating HTML Sanitization Filters: Event Handlers: \url{https://support.portswigger.net/customer/en/portal/articles/2590822-xss-beating-html-sanitization-filters-event-handlers}} However, it does not have general support for detecting mismatches of encoding algorithms and browser contexts. For example, when the user input is URL encoded and ends up in a \texttt{javascript:} scheme URL, Burp does not find it, even though it is exploitable for an arbitrary JavaScript execution.

KameleonFuzz~\cite{10.1145/2557547.2557550} is a blackbox fuzzer of web applications which uses browser parsing for context sensitivity. KameleonFuzz uses model inference for the detection of control flow and taint flow and a genetic algorithm (called \emph{evolutionary fuzzing}) for the generation of exploit payloads. Tripp et al.~\cite{10.1145/2483760.2483776} use a learning approach to blackbox testing. Their solution uses a very large space of attacking payloads and prunes this space using feedback from the failed exploitation attempts. This approach is also context-sensitive because it parses HTML pages and performs syntactic analysis over input points (e.g. HTTP parameters). The blackbox XSS detection approach described in~\cite{10.1007/978-3-319-33630-5_17} proposes new quality metrics for XSS vulnerability scanners. The approach was evaluated on an experimental testbed that contained only reflected XSS flaws and its authors came to a conclusion that existing blackbox XSS scanners have limited context awareness when choosing a potential exploit payload.

\subsection{Whitebox Detection of Context-Sensitive XSS Vulnerabilities}

Among the related work that focuses on systematic context-sensitive XSS detection, but does not use the blackbox approach, we have DjangoChecker~\cite{doi:10.1002/spe.2649}. DjangoChecker is a dynamic whitebox analyzer that instruments the application under analysis and runs on the HTTP server side. Because of the whitebox character, DjangoChecker's implementation is bound to a specific web application runtime. Similarly to our approach it uses recursive parsing of the HTTP response bodies for the detection of web browser contexts.

A whitebox tool with a purpose similar to DjangoChecker is SCRIPTGARD~\cite{Saxena:2011:SAC:2046707.2046776}. SCRIPTGARD aims at dynamically autocorrecting sanitizations in legacy applications. It works in two phases. During the training phase, it computes correct sanitizations and builds a sanitization cache that maps each execution path to sanitization algorithms. During the production phase, if SCRIPTGARD encounters an execution path that is already in the sanitization cache, it applies a matching sanitization. Otherwise, the associated request is either blocked or allowed to proceed unchecked. The training phase can be used as a separate analyzer.

A combination of static and dynamic whitebox analysis is performed by the Context-Sensitive Autosanitization (CSAS) approach~\cite{Samuel:2011:CAW:2046707.2046775}. CSAS works with templating languages that generate HTML output, and first attempts to determine the web browser context of unsafe values statically using type qualifiers. If the static computation succeeds, proper sanitization is hardcoded during compilation of the template. Otherwise, CSAS generates a runtime check, which determines the web browser context dynamically. The CSAS approach is useful for applications developed from scratch, however, it is limited to a class of sufficiently restricted templates. Navex~\cite{alhuzali2018navex} also combines static and dynamic analysis of PHP code in order to automatically generate XSS exploits. It does not check browser contexts, but it validates sanitizer correctness similarly to our work. Navex has a different purpose than our scanner, because it automatically constructs exploits for the identified vulnerabilities. A purely static whitebox approach towards the detection of context-sensitive XSS flaws is JspChecker~\cite{Steinhauser:2016:JSD:2993600.2993606}, which analyzes JSP applications. Similarly to our work it recursively parses HTML outputs in order to compute web browser contexts.

\medskip

The existing approaches towards context-sensitive XSS flaw detection rely on the ability to trace data flows inside the web application. That is why they cannot be used by non-whitebox scanners. Our approach is novel, because it does not have the dataflow analysis requirement but still handles context sensitive XSS (we detect the exploit payload post-hoc using regular expression matching and simulate browser decoding in order to verify the sanitization).

A systematic analysis of XSS sanitization in web application frameworks in~\cite{Weinberger:2011:SAX:2041225.2041237} does not propose any particular approach towards context-sensitive XSS flaw detection. However, it succinctly describes subtleties and challenges in context-sensitive XSS sanitization and statistics of context-sensitive XSS bug-patterns in real-world web applications. Those statistics roughly correspond to our own observations during the evaluation of our approach.

\section{Conclusion}
\label{sec:conclusion}

We have presented an approach to XSS analysis that advances the state-of-the-art by detecting XSS flaws that are stored or context-sensitive or both. The approach relies on moving the database out of the blackbox (a feasible step given that many databases are standalone processes connected to web applications through documented protocols), and coordinates the exploit payload injection between the analyzer and the database. Furthermore, the approach uses flexible exploit payload identification based on regular expression matching, coupled with browser context analysis, to identify even encoded instances of the exploit payload.

Our approach is highly practical -- it was evaluated on eight mature and technologically diverse web applications, and succeeded in identifying highly severe stored context-sensitive XSS flaws in all of them. Furthermore, the analysis was done in reasonable time and with high precision -- out of 79 incorrect sanitization instances flagged by our approach as permitting arbitrary JavaScript execution, we have manually demonstrated exploitability through the web application's user interface in 66 cases. We have further demonstrated that three state-of-the-art blackbox scanners fail to detect these same vulnerabilities, and are in fact not systematically effective against stored and context-sensitive XSS flaws. We have reported a total of 71 manually verified flaws that permit arbitrary JavaScript executions to the application maintainers, some are listed in the CVE database, most are also already resolved.

We are currently implementing a proxy server to wrap mainstream database protocols (MySQL, PostgreSQL, MSSQL and Oracle) with the request reporting and response injection support. This will allow us to encapsulate the database in a wrapper implemented in Java and distributed along with our ZAP plugin, instead of the current solution that modifies an existing database implementation. This will allow us to intercept traffic to non open-source databases (Microsoft SQL Server or Oracle Database).

We see perspective future work in the implementation of dataflow analysis for JavaScript string content~\cite{Kashyap:2014:JSA:2635868.2635904,Gauthier:2018:FRD:3236454.3236502,7476648}. Connecting our current technique together with the dataflow analysis will lead to systematic detection of a new class of XSS flaws which might be widespread in popular applications. Detection of JavaScript-embedded languages can also be connected with DOM XSS detection~\cite{184491,Parameshwaran:2015:DRT:2786805.2803191}.

\section*{Acknowledgments}

The work on this paper was partially sponsored by Charles University institutional funding SVV 260451 and 260588.

{\normalsize \bibliographystyle{acm}
\bibliography{biblio.bib}}

\end{document}